\documentclass[manuscript,nonacm]{acmart}

\usepackage[english]{babel}
\usepackage{csquotes}
\usepackage{parskip}

\AtBeginDocument{%
  }

\setcopyright{acmlicensed}
\copyrightyear{2019}
\acmYear{2019}
\acmDOI{XXXXXXX.XXXXXXX}

\acmJournal{JACM}
\acmVolume{37}
\acmNumber{4}
\acmArticle{111}
\acmMonth{8}

\newcommand\region[1]{{\textcolor{blue}{#1}}}

\newcommand{\comm}[1]{}

\begin{document}
\graphicspath{ {./images} }

\title{Censorship Chokepoints: New Battlegrounds for Regional Surveillance, Censorship and Influence on the Internet}

\author{Yong Zhang}
\affiliation{%
  \institution{University of Surrey}
  \city{Guildford}
  \state{Surrey}
  \country{United Kingdom}
}

\author{Nishanth Sastry}
\affiliation{%
  \institution{University of Surrey}
  \city{Guildford}
  \state{Surrey}
  \country{United Kingdom}
}


\begin{abstract}
 \section{Abstract}
 
 Undoubtedly, the Internet has become one of the most important conduits to information for the general public. 
 Nonetheless, Internet access can be and has been limited systematically or blocked completely during political events in numerous countries and regions by various censorship mechanisms. Depending on where the core filtering component is situated, censorship techniques have been classified as client-based, server-based, or network-based. However, as the Internet evolves rapidly, new and sophisticated censorship techniques have emerged, which involve techniques that cut across locations and involve new forms of hurdles to information access. We argue that modern censorship can be better understood through a new lens that we term \textit{chokepoints}, which identifies bottlenecks in the content production or delivery cycle where efficient new forms of large-scale client-side surveillance and filtering mechanisms have emerged. \comm{We argue that many modern mechanisms go beyond the classical censorship mechanisms which involved hard and noticeable restrictions to information access. Instead, these soft chokepoints may redirect users' attention, manipulate search results or modify AI models, thereby making it much harder to access certain kinds of information.}
\end{abstract}



\maketitle

\section{Introduction}

Internet censorship and surveillance involves observing, controlling, subverting or suppressing the publication of or access to online information. Governments, institutions or private parties may engage in censorship to restrict access to content they find objectionable, threatening, or contrary to their interests. It has become de rigueur to exercise surveillance in politically turbulent scenarios. For instance, \region{Iranian}\footnote{Country names are coloured in blue throughout the paper to enable readers to find regional examples easily.} authorities have restricted content that they regard as a danger to their power or stability, such as news stories, social media posts, and websites discussing political problems or criticising the administration~\cite{aryan_internet_2013}. Even stable democracies may resort to censorship. For instance, censorship activities were observed in \region{Spain} during the Catalan independence referendum in 2017~\cite{ververis_understanding_2021}. Similarly, \region{India} was classified as only ``partly free'' by the 2024 Freedom on the Net report~\cite{noauthor_freedom---net-2024-digital-booklet_2024}. 
We have also seen commercial platforms restricting content or users (e.g., Twitter's 2021 Trump ban~\cite{xcom_permanent_2021}); governments imposing bans on social media platforms, as seen in \region{Nepal}~\cite{shamim_nepal_2025} and \region{Indonesia}~\cite{safenet_statement_2025}; or even worse, authorities shutting down the Internet entirely, as occurred in \region{Afghanistan} recently~\cite{netblocks_afghanistan_2025}.

\begin{figure*}[htb]
   \includegraphics[width=1.0\linewidth]{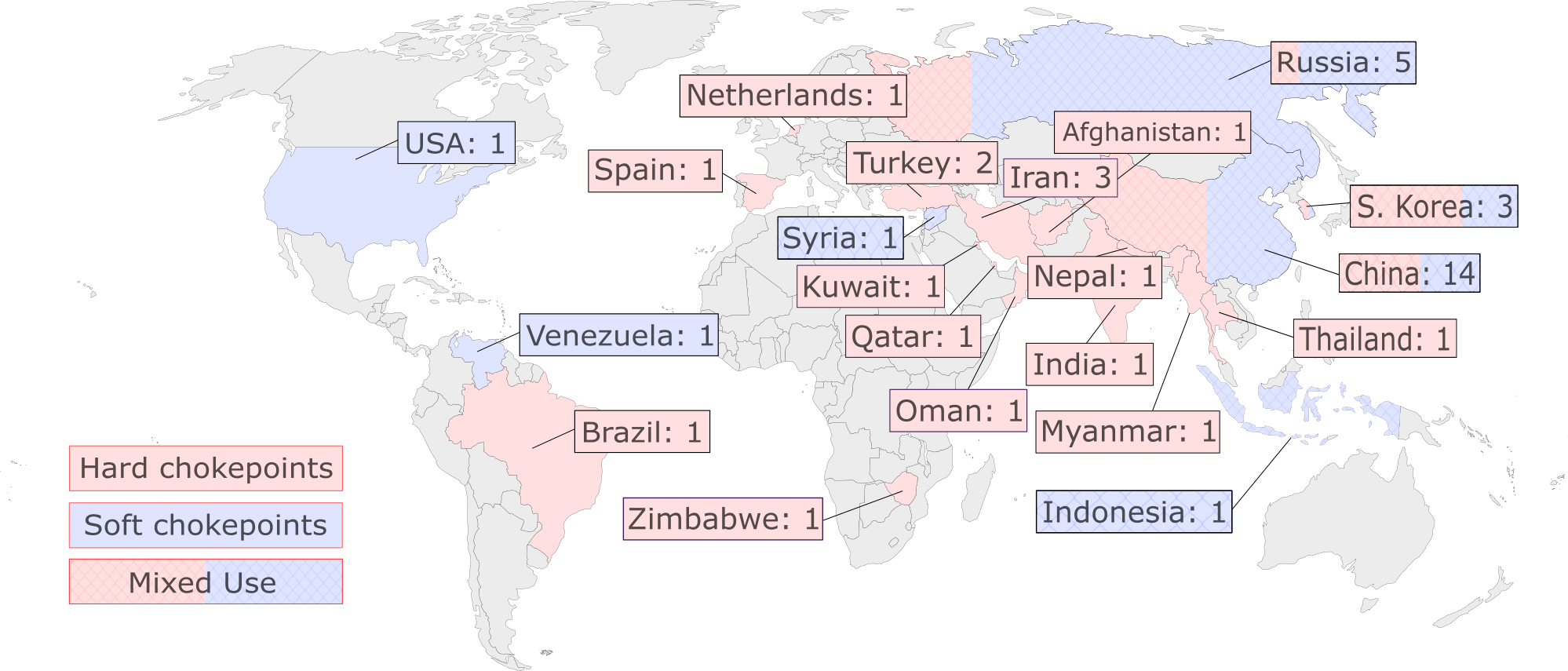}
    \caption{Geographic distribution of censorship examples used in this study. Numbers indicate numbers of case studies from each country. Red (blue) indicates `hard' (`soft') chokepoints. Mixed use combines both colours.}
    \Description{Geographic distribution of our examples}
    \label{fig:geo-distribution}
\end{figure*}

The goal of this paper is to provide an overview of the current state of Internet censorship and how it has been changing in recent years. At the outset, it is useful to state explicitly that we wish to stay value neutral and it may not be possible to take a strict moral stance that censorship is always `wrong' --- there may be valid geopolitical imperatives such as peace keeping due to which a sovereign state or government may feel impelled to implement censorship or surveillance, e.g., in order to minimise the spread of disinformation in politically fraught circumstances. \comm{Governments across the world have seen it necessary (or convenient) to deploy censorship. }\autoref{fig:geo-distribution}, which depicts just the examples used in this paper, is an indication of how widely censorship is practiced.\comm{This paper does not take political positions, but focuses on how censorship and countermeasures are evolving, especially in strained political circumstances, and the research opportunities presented.}

Given the ever-rising prevalence and effects of Internet censorship, several prior surveys of Internet censorship exist. They have each developed a different way of looking at the existing censorship practices at the time. For example, in 2010 the Open Net Initiative (ONI) classified censorship into four time-based phases, starting with the ``Open'' Internet up until 2000, followed by an ``Access Denied'' stage from 2000--05, when filters were deployed, after which there was an ``Access Controlled'' stage when a chilling effect~\cite{alex_marthews_government_2017} developed among Internet users in censored locations, which later led to an ``Access Contested'' stage (as of 2010) where a conversation was started about the extent to which control should be allowed or exercised~\cite{palfrey_four_2010}. As we will argue later, we may have moved to a new stage where access control is being exercised covertly, thereby making it difficult to even realise that censorship is occurring, which is a key first step to enabling the contested conversation. Orthogonal to this time-based taxonomy, Aceto and Pescapé~\cite{aceto_internet_2015} categorise censorship based on the main locations where the censorship takes place: client, network or server. 

This paper is an attempt to update such previous scholarship on censorship to take into account new censorship forms that are emerging. For instance, YY\footnote{https://en.wikipedia.org/wiki/YY.com}, a major \region{Chinese} video-based social network incorporates client-side censorship mechanisms within their chat functions, and in addition, if a chat contains a keyword from one of three (regularly updated) blacklist files, the app's surveillance feature is triggered, resulting in the censored content, along with both the sender and receiver's user IDs, being transmitted to the remote server~\cite{knockel_every_2015}.  Such censorship, combining client and server methods, cuts across multiple modalities of Aceto and Pescap\'e's taxonomy.\comm{From a user's perspective, although information expression may not always be restricted by the platform, just the knowledge of the existence of surveillance operations can create a form of self-censorship.}\comm{, wherein users hesitate to express controversial views.}

Recent years have also witnessed the rise of new forms of information control which do not suppress or restrict access to information, but instead direct the \textit{attention} of users in ways that may serve the purpose of an important actor such as a state.
\comm{Recent years have also witnessed the rise of new forms of information control in which the ways access is restricted go beyond the ONI taxonomy (access denied vs.\ controlled vs.\ contested). These new censorship practices do not suppress or restrict access to information, but instead direct the \textit{attention} of users in ways that may serve the purpose of an important actor such as a state.}For instance ``attention honeypots'' draw users' attention into some unconnected content and thereby implicitly silence genuine content during political events. For example, before and during the 2016 US presidential election, \region{Russia}-sponsored  troll accounts with elaborate Twitter (X) profiles are believed to have been employed in the USA to spread disinformation or drive traffic to specific webpages, diverting users' attention from genuine political content~\cite{zannettou_disinformation_2019}. 
\begin{figure*}
\includegraphics[scale=0.35,width=\linewidth]{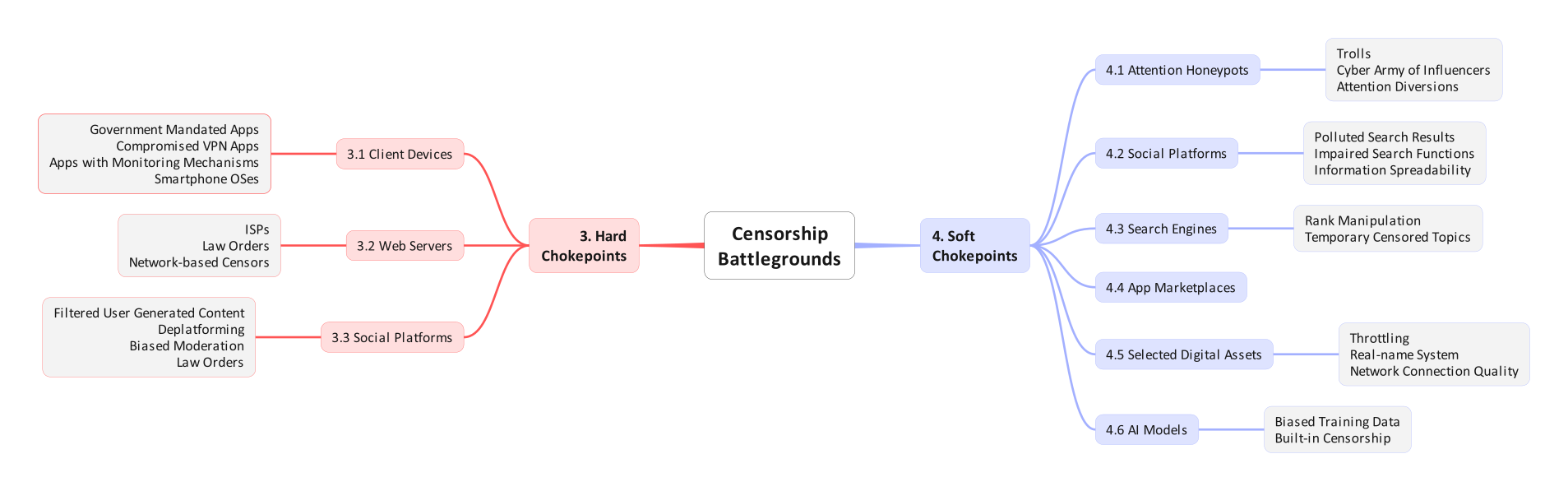}
    \caption{A new Taxonomy of Internet Censorship based on Soft and Hard Chokepoints}
    \Description{Censorship Soft Chokepoints and Hard Chokepoints}
    \vspace{-1em}
    \label{fig:censorship-chokepoints}
\end{figure*}

Unlike traditional forms of Internet censorship, these new tactics do not by themselves restrict (deny or control or contest) access to information in well defined network locations, yet they serve similar aims: \comm{Yet, the new mechanisms still aim to achieve similar outcomes as traditional censorship: }to suppress `inconvenient' views or influence opinion towards more `convenient' ones. Thus, while previous studies have offered valuable insights into Internet censorship practices, this paper seeks to advance our understanding of modern Internet censorship by introducing a novel taxonomy (\autoref{fig:censorship-chokepoints}) that aims to reflect rising new forms of Internet censorship.\comm{ Our study can be seen as elaborating on ONI's phase four~\cite{palfrey_four_2010}, in which Internet access is increasingly challenged. }

We can think about modern censorship in terms of \textbf{chokepoints} --- critical places where high throughput information relay occurs, or in the view of censorship systems, where digital information control can be most effectively exercised. Modern censoring infrastructure can run information control against a particular chokepoint (e.g. mobile devices) or multiple chokepoints simultaneously as seen in the previous YY.com example~\cite{knockel_every_2015} \comm{which involves client and server-side surveillance}, covering the complete data flow journey from the user to the app, and then to the server. We can recognise two kinds of chokepoints: \textit{hard chokepoints} and \textit{soft chokepoints}.
\textbf{Hard chokepoints} are where explicit and permanent censoring occurs, causing unrecoverable information loss. Examples of hard chokepoints include social platforms in authoritarian regimes, where criticism posts or activists' accounts could be permanently removed. Such information suppression can also occur on client devices~\cite{knockel_measuring_2021}, web servers~\cite{hoang_measuring_2019}, AI models~\cite{noels_what_2025} or app marketplaces~\cite{ververis_shedding_2019}.
\textbf{Soft chokepoints} restrict or control information relay points in an implicit and less noticeable manner without causing permanent information loss. Soft chokepoints could involve slowing down network flows or otherwise affecting connection quality~\cite{aceto_internet_2015}, attention honeypots that divert users' attention without them noticing~\cite{zannettou_disinformation_2019}, search engines whose result rankings are silently manipulated~\cite{jiang_business_2014}, or AI models whose answers are aligned to a government's interests~\cite{hoang_empirical_2018}.

While hard chokepoints typically require the owner of an Internet platform or resource to exercise censorship, soft chokepoints can be exploited by trolls or a cyber army of influencers who may not have full control of a platform, as seen in various elections \region{worldwide}~\cite{zannettou_disinformation_2019}.  \comm{Because soft chokepoints do not directly cause information loss, it is possible for committed users to access the information. However, they can be highly effective because their additional technological and cognitive burden can discourage a majority of users. }Both hard and soft chokepoints can work across multiple network locations of the Aceto-Pescapé taxonomy (e.g., across client and server in the YY example above), and may commonly go beyond the ONI's `Access Contested' stage of censorship, since many users may not realise their information access is being manipulated (e.g., with attention honeypots) or users may not have a choice in the matter (e.g., when client apps censor based on keywords or other rules). 




We summarise the main differences between the characterstics of hard and soft chokepoints in Table \ref{tab:chokepoints-comparison}. The primary distinction between hard and soft chokepoints lies in their implementation and detectability. Filtering at soft chokepoints often resembles genuine network activities, making it implicit and hard to detect, and it often does not cause permanent information loss; therefore, users' reactions can be expected to be milder. In contrast, filtering at hard chokepoints tends to be more direct and permanent and often causes unrecoverable information loss, making it easily noticeable and always follows agitated user reactions because of the impact and disruption, as seen after an anti-fraud app was installed on millions of mobile devices~\cite{xue_anti-fraud_2021}.

\begin{table}[ht]
    \centering
    \caption{Hard Chokepoints vs Soft Chokepoints}           
    \begin{tabular}{p{0.3\linewidth} p{0.32\linewidth} p{0.32\linewidth}}
        & Hard Chokepoints & Soft Chokepoints \\
    \hline
        Noticeability & Obvious & Subtle \\
        Filtering methods & Explicit & Implicit \\
        Impact & Direct & Indirect \\
        Damage & Permanent & Temporary \\ 
        Collateral damage & Severe & Mid-level \\  
        Detectability & Easy & Hard \\
        Cost to implement & Medium/High & Low/Medium \\
        Usage frequency & Occasionally used & Frequently used \\
        Countermeasures & Limited & Easy to find \\
        Example & Client devices & Attention honeypots \\
    \hline
    \end{tabular}
    \label{tab:chokepoints-comparison}
\end{table}

In summary, our main contributions are fivefold:
\begin{enumerate}

\item To the best of our knowledge, our study is the first comprehensive attempt to define and categorize censorship practices from across the world (\autoref{fig:geo-distribution}) as hard and soft chokepoints  (\autoref{fig:censorship-chokepoints}) while simultaneously bridging the gap between censorship and resistance mechanisms. 
\item We identify and incorporate new forms of censorship into our taxonomy, highlighting the rise of soft chokepoints and their potential for exercising new powerful forms of information control.
\item We examine recent client-side censorship and surveillance practices in countries where these methods are most prevalent, highlighting a significant increase in large-scale occurrences.
\item We emphasise the need for greater scrutiny of ML-powered filtering technologies from researchers and the general public, given their growing influence on censorship.
\item We review new countermeasures against both soft and hard chokepoints, including the recent development of decentralised networks and innovative strategies to bypass censorship.
\end{enumerate}


\section{Hard Chokepoints} \label{hardcensorship}
In response to political events or online user activities, some governments or regimes may implement censorship or surveillance mechanisms at hard chokepoints to restrict the development of the event by eliminating sensitive information efficiently and permanently. Hard chokepoints are difficult for lay users to circumvent, but are also more easily noticed. 

\subsection{Client-based Surveillance Support Systems}
While client-side censorship is less widely recognised than server-based and network-based filtering techniques, we argue that this mechanism has been increasingly implemented in modern systems as an efficient and cost-effective method for information control, especially via the threat of surveillance, which may cause a chilling effect on ways client devices are used. These mechanisms are typically billed as anti-fraud measures, but the data they collect may also potentially be used to surveil and identify users. Such measures can be embedded into apps, some of which may be mandated by authorities, or into Phone Operating Systems (OSes). This form of covert censorship can be technically difficult to bypass since the filtering components can operate in the background. \comm{Client device-based censorship can be extremely effective because of the following reasons:

\begin{description}
\item[Ahead of time circumvention] Client-side filtering mechanisms being on the device allow censorship components to get ahead of any end-to-end encryption (E2EE) that may prevent more sophisticated analysis in the network. This reduces public exposure to censorship events and lowers manual content moderation costs.

\item[Ubiquity of censorship] The pervasive nature of client-side filtering may induce a so-called ``chilling effect''~\cite{alex_marthews_government_2017}, leading to increased user self-censorship or acceptance of censorship practices. 

\item[Implementation efficiency] Client-side censorship often leverages lightweight keyword-matching or rule-based approaches which are easy to implement.

\item[Difficulty of circumvention] Ordinary users may find it challenging to bypass client-side filtering embedded in software or apps which are in binary format, unlike network-based censorship, which could be circumvented simply by using VPN tools.
\end{description}}

In 2015, \region{South Korea} mandated telecommunications companies to install a censorship app on all mobile phones sold to individuals under 19 to prevent access to online content deemed objectionable or harmful~\cite{borowiec_south_2015}. Similarly, in 2021, \region{China} launched an anti-fraud app developed by China's Public Security Ministry's National Anti-Fraud Centre. The app’s primary goal is to warn users about potential cyber fraud and report suspicious phone calls and online communication to the police. It claims to have a database of 3.8 million
fraud-related websites and half a million apps. However, the app can access users' live phone logs and active cellular data, including cellular base station registration information. While not mandatory, some local governments pressured citizens and students to install it, resulting in its ranking once among China's top 10 most downloaded apps~\cite{xue_anti-fraud_2021}. Concerns have been raised regarding the extensive personal data collected during registration, including real names, ID numbers and phone numbers~\cite{xue_anti-fraud_2021}, which could be used for surveillance. 

Nevertheless, private software companies may be also embedding censorship or surveillance code, often due to government pressure. Rodriguez \textsl{et al.}~\cite{rodriguez_revisiting_2025} recently analysed three \region{Chinese} Android browsers namely UC Browser, QQ Browser, Baidu Searchbox, along with three browsers pre-installed on the OPPO, Xiaomi and VIVO phones (important players in the Chinese market) and showed that sensitive data such as PII, geolocation and browsing activity data are uploaded to the browser vendors' servers, unencrypted or poorly encrypted using symmetric algorithms, even in `incognito' mode.  Liu \textsl{et al.}~\cite{liu_android_2021} revealed that Samsung, Xiaomi, Huawei, and Realme phones constantly gather user data through system apps preinstalled in protected read-only disk partitions. These systems apps have substantial permissions and can transmit user activity data to remote servers without a clean `opt out' choice. Device identifiers and IMEI are among the data collected. Knockel \textsl{et al.}~\cite{knockel_measuring_2017} also found that client-side keyword censorship was common in popular \region{Chinese} mobile games, with game publishers largely responsible for maintaining lists of prohibited keywords. However, it remains uncertain whether these lists originate directly from the government.

\subsection{Web Servers} Some web servers function as critical proxies or nodes of widely utilised censorship countermeasure infrastructures, such as the Tor network or The Invisible Internet Project (I2P). Consequently, they have become the primary targets of censorship in a few countries to increase the difficulty and cost of accessing information freely. Hoang \textsl{et al.}~\cite{hoang_measuring_2019} recently uncovered active censorship of these server nodes in \region{China, Iran, Oman, Qatar, and Kuwait} using various methods such as domain name blocking, TCP packet injection, or injecting pages (through mechanisms such as DNS poisoning) declaring that access to a web resource is blocked.

A common and now classical means of censorship by governments is to use court orders and warrants to order Internet Service Providers (ISPs) to block websites deemed objectionable. Ververis \textsl{et al.}~\cite{ververis_understanding_2021} found that some Spanish ISPs employed techniques such as HTTP blocking, Deep Packet Inspection and DNS poisoning to restrict access to specific sites during the Catalan referendum, and that domains related to the Catalan independence referendum were seized under court warrants. In \region{India}, several thousand websites have been blocked by various ISPs utilising filtering techniques such as DNS tampering, HTTP DoS, and SNI (Server Name Indication), as a result of legal orders from the central government or courts~\cite{singh_how_2020}. However, the censorship practices differ among ISPs; for instance, SNI Inspection was only implemented by one ISP, Jio, according to the study.

In addition to government-enforced censorship, websites may implement internal filtering to comply with local laws and regulations. For instance, recent research~\cite{knockel_measuring_2021} indicates that certain keyword combinations can indiscriminately trigger Tencent's QQMail automated censorship system, regardless of whether the \textbf{Mail Exchange (MX)} server is based in \region{Hong Kong} or \region{Mainland China}~\cite{knockel_measuring_2021}. 

\subsection{Verifiable IDs}
In 2007, \region{Korea} became the first country to enforce real-name systems on certain websites, meaning any online activity is now traceable to an individual. These systems may reduce indiscreet online posting behaviors to some degree, it could also be an obstacle to freedom of speech~\cite{schneier_real_2013}. Similarly, in 2013, SIM card registration became mandatory in \region{Zimbabwe}~\cite{yingi_future_2025}, and \region{China} is also moving in this direction, and may soon require every Internet user to have a unique Internet ID~\cite{tobin_china_2024}. This goes beyond real names and can identify individual users easily. 
On the server side, any website operating from \region{China} must obtain an Internet Content Provider license~\cite{zhen_lu_accessing_2017} according to relevant regulations, imposing accountability on the server side as well. Such measures can lead to self-censorship, as users may hesitate to express themselves online because of the threat of punitive measures if the regime manages to identify them using a Verifiable ID.

\subsection{Social Platforms}
\label{sec:social_hardchoke}
Given their immense popularity, social platforms are an important target for censorship. For instance, a court order in \region{Brazil} resulted in the blocking of X~\cite{kahn_x_2024}, as the company declined to appoint a legal representative. This decision left the platform's largest user base, consisting of approximately 21 million users, disconnected from a critical source of information~\cite{kahn_x_2024}. Nevertheless, some argue that whether the ban is an attack on freedom of expression is yet to be studied since Brazilian users can still access other social platforms. In fact, over 2.5 million users migrated to BlueSky\footnote{https://bsky.app} in just a few days following the ban~\cite{kahn_x_2024}. In countries such as \region{Myanmar} and \region{Zimbabwe}, social networks have been censored or completely blocked during riots~\cite{padmanabhan_multi-perspective_2021, yingi_future_2025}. 
However, it is more common for social platforms to implement partial censorship measures by filtering content or deplatforming (e.g. suspending accounts) that conflict with their policies or interests. In \region{Turkey}, since the Internet censorship law was passed in 2007 with the aim of a cleaner online environment, more and more websites have been removed or banned, especially after the Gezi events in 2013~\cite{ververis_cross-country_2020}. As a result, social media websites such as YouTube, Facebook, and Twitter were constantly or intermittently blocked~\cite{ververis_cross-country_2020}. Haimson \textsl{et al.}~\cite{haimson_disproportionate_2021} argue that \region{globally}, marginalised groups, such as political conservatives, trans people and black individuals, may experience disproportionately more social media content or account deletions, making it even more challenging for their voices to be heard.

Nonetheless, toxic content and hate speech have also become substantial challenges recently. As a countermeasure, `deplatforming'~\cite{rogers_deplatforming_2020}, is employed by social platforms as a moderation method to maintain civil discourse. This involves denying a user or group of users access to the platform, e.g., based on their previous activity, such as posting hateful or inciteful content. However, `deplatforming' can also be utilised to silence voices in repressive regimes~\cite{knockel_every_2015}. 
Social media platforms often suspend or permanently block user accounts believed to promote criticism or collective actions to comply with government orders. 
Biased moderation, whether automated or manual, may have the same effect as censorship. 

We classify these as `hard' chokepoints because users who are aware of these forms of censorship may not be able to get around them  (e.g., Brazil users moving to BlueSky when Twitter/X was blocked). However, these chokepoints also share some aspects of soft chokepoints since some users may not be aware that information is being censored, or indeed, \textit{what} is being censored. For instance, although users may be aware that certain servers or even apps on their devices may have monitoring  mechanisms such as searching for sensitive keywords, the precise list of keywords may not be public, so the users may resort to overly conservative behaviour, avoiding discussion of anything that might be even remotely construed as controversial.


\section{Soft Chokepoints} \label{softcensorship}
Unlike hard chokepoints, filtering at soft chokepoints aims to imitate genuine network activities so that access to sensitive content is restricted indirectly and implicitly. By conducting censorship at soft chokepoints, the dissemination of sensitive information may slow down, and the restriction is less noticeable, giving autocratic regimes additional time to address public challenges. We have observed the following soft chokepoints in literature. 

\subsection{Attention Honeypots}
During an ongoing public affair, autocratic governments often find it difficult to effectively block all criticism posts appearing on social media platforms when they are being blamed. In these situations, the government may recruit a covert cyber army of hundreds or even thousands of real accounts to post pro-government comments on news websites, forums and chatrooms to influence public opinions~\cite{king_how_2013} (\region{China}) or divert users attention away from trending topics they dislike~\cite{zannettou_disinformation_2019} (\region{Russia, in USA}). In cases where the attacker has limited control over the servers, a data poisoning attack could also be executed to disrupt the platform's search functions. Muñoz \textsl{et al.}~\cite{munoz_modeling_2024} conducted a study on \region{Spanish} Twitter accounts and content, revealing that the disinformation botnets consisting of 275 bot accounts were more active and generate more content than legitimate journalists, making it difficult, if not impossible, for information seekers to find legitimate information on specific topics. For example, their study~\cite{munoz_modeling_2024} shows the botnets' efficiency spiked when the Russian-Ukraine war broke out. 
Note that the same tactics can also be used by non-government actors, including private individuals acting as trolls, or hostile governments seeking to influence opinion during sensitive periods such as elections\cite{zannettou_disinformation_2019}.

\subsection{Social Platforms}
Due to their importance in modern online discourse, we find not only hard chokepoints (\autoref{sec:social_hardchoke}) but also soft chokepoints. These typically take the form of so-called `shadow' or `stealth' banning. This involves suppressing specific comments and posts related to some topics, or all activities of selected users. For instance, Ryan \textsl{et al.}~\cite{ryan_tiktok_2020} reported that TikTok was \region{globally} found to suppress LGBTQ+ related hashtags in more than eight languages and specific hashtag search results during political events such as the George Floyd protests. Another study~\cite{thomas_studying_2021} on Internet outages in \region{Venezuela} in January 2019 by Thomas \textsl{et al.} reveals that some users could serve as information gatekeepers who can influence information circulation speed during political events. These examples demonstrate the critical role social platforms play as battlegrounds. Risius and Blasiak~\cite{risius_shadowbanning_2024} argue that four types of shadowbans can be identified on social platforms based on content visibility levels: Ghost bans (content visible only to the creator), Search bans (content removed from search results), Search suggestion bans (content excluded from search suggestions), and Downtiering (content visibility reduced algorithmically). Their study suggests that shadowbanning may also lead to biased content moderation~\cite{risius_shadowbanning_2024}. 

\subsection{Search Engines} The search engine is another major chokepoint constantly targeted by authoritarian regimes due to its tremendous influence on our activities. This surface can be and attacked from inside or outside, utilising proactive filtering, manipulation of search results or rankings, and Search Engine Optimisation (SEO) poisoning. For example, Jiang's comparative study~\cite{jiang_business_2014} of Google and Baidu highlights Baidu's filtering on \region{Chinese} search results is more fine-tuned, and the links to an external Chinese Wikipedia, Hudong Baike, are often hidden. Jiang~\cite{jiang_business_2014} demonstrates that Baidu's search results have higher accessibility (the availability of a given search result) than Google, to be specific, out of 3160 search query results tested, Baidu only had 171 inaccessible links (which are banned in China) whereas Google had 400. Jiang~\cite{jiang_business_2014} argues that this could be a mixed result of firm's system algorithms, self-censorship and the government policies. The higher availability of Baidu's search results could make its bias and filtering less noticeable and lead to less stronger user reactions; hence we classify manipulation of search engine results as a soft chokepoint. 

\subsection{App Marketplaces}
Accessing information, such as news and trending stories, through smartphones, tablets, and other smart devices has become a daily routine for many people. With millions of apps available on the AppStore, Google Play and other third-party marketplaces, it is unsurprising to see apps removed due to quality issues, intellectual property infringement, and other reasons. Nevertheless, apps could also get delisted for political reasons. A recent study by Ververis \textsl{et al.}~\cite{ververis_shedding_2019} finds that app censorship varies by app and country due to local laws and regulations. For instance, the research shows that even though VPNs are banned in \region{Russia}, a keyword search for `VPN' still returned nearly 200 records. At the same time, only a few government-approved and closely monitored VPN apps are available in the \region{Chinese} App Stores. This indicates that much thorough app censorship has been applied across digital app marketplaces in China.

\subsection{AI Models}
AI technology has advanced immensely and has been adopted by lots of apps we use every day. AI-driven chatbots have become extremely useful and popular, and therefore not immune to scrutiny by censors, particularly when handling sensitive topics. Joe Yizhou Xu examined two popular AI chatbots~\cite{xu_programmatic_2018}, developed by Microsoft Research China and Tencent, revealing that both chatbots were aware of political topics unsuitable for public discussion and tried to avoid answering users' sensitive inquiries.
Just lately, Noels \textsl{et al.}~\cite{noels_what_2025} reveals that some LLMs have embedded censorship patterns closely linked to their geopolitical origins. For example, \region{Chinese} LLMs such as Deepseek and Qwen are highly likely to reject queries about Chinese political figures, while \region{Russian} LLMs GigaChat and YandexGPT tend to refuse many questions related to Russian-born figures. 

An even more insidious form of censorship experience may result as an implicit side effect of training on datasets that may themselves have been censored or moderated. Websites, books, articles or even online encyclopedia corpora, such as Wikipedia corpora, are often used to train word embeddings due to their extensive and consistently updated high-quality data. However, word embeddings trained on censored or moderated corpora could introduce biases into predictive models. For example, comparing the responses generated by ChatGPT with those generated by humans, Vahid Ghafouri \textsl{et al}~\cite{ghafouri_ai_2023} demonstrated that it is difficult to completely mitigate bias from AI responses due to moderation or biased training data, despite modern AI systems having been continuously improving, aiming to be neutral when asked about controversial topics.
In another research, Yang and Roberts~\cite{yang_censorship_2021} argue that word embeddings trained on Baidu Baike, an alternative to the blocked \region{Chinese} language Wikipedia, exhibit less association with democracy, freedom, government monitoring and historical events. Baidu Baike's corpora have been increasingly used for making Chinese apps powered by NLP~\cite{yang_censorship_2021}. As a result, Chinese users may encounter more implicit censorship if the apps they use employ language models trained upon Baidu Baike's word embeddings.

\subsection{Network Throttling and Monitoring}\label{sec:throttling}
During political events, individual websites could be intentionally throttled to slow the spread of information. For example, Twitter's accessibility was throttled nationally by the \region{Russian} government in March 2021 to coerce the social media firm to take down content published by the activists~\cite{xue_throttling_2021-1}. The filtering mechanism working at this chokepoint can also disrupt the retrieval of digital assets~\cite{aceto_internet_2015} by worsening the network \textbf{Quality of Service (QoS)}, causing symptoms of loss of QoS, including increased packet loss and jitter, network delays, and throttling in Internet connection. For instance, \region{Iran} was found to disrupt Internet connectivity by throttling Internet protocols and speeds and sometimes all traffic along with other censorship mechanisms, such as Deep Packet Inspection, before the June 2013 presidential election~\cite{aryan_internet_2013}. 

While often used to protect privacy, retain anonymity or circumvent censorship, some VPN software has also been observed intercepting or manipulating traffic, or even disrupting its anti-censorship functions. Khan \textsl{et al.}~\cite{khan_empirical_2018} observed instances of URL redirection and outgoing traffic blocking based on country-specific censorship in \region{Turkey}, \region{Russia}, \region{South Korea}, \region{the Netherlands}, and \region{Thailand}.

\section{Censorship Countermeasures}
Early research on censorship resistance systems focused on designing and implementing the resistance of publishing systems against censorship or inventing new covert channels to access censored content. Khattak \textsl{et al.}~\cite{khattak_sok_2016} argue that most censorship resistance systems are vulnerable to denial-of-service attacks or information poisoning. However, as censorship becomes more ubiquitous and sophisticated, modern censorship resistance techniques have also advanced simultaneously towards stronger and more robust systems, such as decentralised and Web3-based systems (blockchain and crypto-based)~\cite{eldem_decentralisation_2025}. 

There are several countermeasures users could take to reduce or even altogether remove the impact of censorship or surveillance carried out on hard (\autoref{tab:resistance_hard_chokepoints}) and soft chokepoints (\autoref{tab:resistance_soft_chokepoints}).

\begin{table*}[htbp]
    \centering
    \caption{Resistance Mechanisms against Hard Chokepoints}
    \begin{tabular}{p{0.26\linewidth} p{0.44\linewidth} p{0.10\linewidth} p{0.10\linewidth}}
      Chokepoints   &  Resistance Mechanisms &  Complexity & Effectiveness \\
    \hline
      Client-based surveillance     & Restrict app privileges in settings  & Low & Medium \\
      support systems               & Use homonyms, misspelt words in messages & Low & Medium \\
                                    & Install a privacy-friendly mobile OS  & Hard & High \\  
      Web servers                   & Host contents on a decentralised network & Hard & High \\
      Social platforms              & Use decentralised platforms (e.g., Bluesky) & Medium & High \\ 
    \hline
    \end{tabular}
    \label{tab:resistance_hard_chokepoints}
\end{table*}

\begin{table*}[htbp]
    \centering
    \caption{Resistance Mechanisms against Soft Chokepoints}
    \begin{tabular}{p{0.26\linewidth} p{0.44\linewidth} p{0.10\linewidth} p{0.10\linewidth}}
    Chokepoints   &  Resistance Mechanisms & Complexity & Effectiveness \\
    \hline
      Attention honeypots               & Use ML to identify state-sponsored trolls & Hard & Medium \\
      Platform's searchability          & Flag inaccurate hashtags using ML or NLP & Medium & Medium \\
      Search engines                    & Use local LLM to retrieve information & Hard & Medium \\
      App marketplaces                  & Install apps from trusted  developers & Medium & Low \\
      AI models                         & Run models locally and reconstruct prompts & Hard & Medium \\
      Individual websites               & Host websites on cloud, using dynamic DNS, publish content on a peer-to-peer network & Medium & Medium \\                  
      Network throttling                & Use satellite technology, such as Starlink. & Medium & High \\
      and monitoring                    & Use security audited VPN clients. & Medium & High \\ 
    \hline
    \end{tabular}
    \label{tab:resistance_soft_chokepoints}
\end{table*}

\subsection{Hard Chokepoints Resistance Mechanisms}

\subsubsection{Client-based surveillance} 

\textbf{Restrict app privileges or use privacy-friendly OSes.} If users suspect pre-installed apps, especially government-sponsored ones, are collecting user data, they may be able to restrict app privileges in their device's system settings or even completely disable the privacy data access or collection permission. Fixing OS-level privacy collection behaviours observed on phone OSes (especially if closed source) is challenging. However, a few privacy-friendly Android-based mobile OSes, usually open-sourced, are now available, such as LineageOS\footnote{https://lineageos.org/} or GrapheneOS\footnote{https://grapheneos.org/}.
\textbf{Use homonyms or misspelt words.} For apps with built-in sensitive keyword filtering mechanisms, using homonyms or misspelt words in messages can bypass such censorship to some extent. For instance, Hiruncharoenvate \textsl{et al.}~\cite{hiruncharoenvate_algorithmically_2021} argue that using homophone-based text substitution can bypass text filtering observed on Sina Weibo (a \region{Chinese} microblogging website). Their approach works because almost every Chinese character has some characters that phonologically sound the same (including tones) or similar; thus, sensitive words become harder for algorithms to detect while still readable by Chinese users. 

\subsubsection{Web servers}\label{sec:hard-res-webservers}

Ross Anderson proposed the eternity service~\cite{anderson_eternity_1996} to guarantee the availability of digitally published work for a particular length of time by replicating data across a network of servers. The content cannot be deleted even by its owner once posted. 
In recent years, mirrored content or distributed web servers have been widely used to guarantee availability. With cloud technology, it takes only a few moments to launch a mirrored site at a low cost or even free with a free tier offered by many cloud service providers. Nonetheless, mirroring does not necessarily move the content away from the censor's sphere of influence (e.g., a new mirror can be blocked by an ISP simply by updating its list of prohibited sites); hence, it usually only works for a short-term. In recent years, the InterPlanetary File System (IPFS), which utilises content-based addressing, has been widely used as a fundamental layer in many decentralised applications to tackle digital censorship. However, a study~\cite{sridhar_content_2023} demonstrates that a Sybil attack can be operated by generating enough malicious peers around a legitimate content provider node to prevent its content from being retrieved. 

\subsubsection{Social platforms} Recent advances in decentralised social platforms demonstrate the increasing public interest in censorship-resistant publishing systems. For example, Mastodon has recently gained extensive popularity after the privatisation of Twitter. Nevertheless, Raman \textsl{et al.} argue~\cite{raman_challenges_2019-1} that decentralised software such as Mastodon may still face censorship challenges as their study indicates almost half of the users registered on only 10\% of total Mastodon instances. The emergence of Nostr (Notes and Other Stuff Transmitted by Relays), an open-source protocol, which allows users to deploy decentralised relays to offer censorship-resistant data storage and transmission, could be an answer to the aforementioned challenges of Mastodon system.


\subsection{Soft Chokepoints Resistance Mechanisms}
\subsubsection{Attention honeypots} Saleh Alhazbi~\cite{alhazbi_behavior-based_2020} argues that classification models trained by ML algorithms can separate normal troll behaviours from state-sponsored ones. The author's study demonstrates that troll accounts on social platforms commonly have different activity patterns even though they try to mimic genuine users: Troll accounts are coordinated by one or more master accounts, a fact which can often expose their suspicious behaviours.

\subsubsection{Social platforms} Recent research~\cite{abeywardana_hashtag_2018} shows it is possible to create hashtags based on features extracted from images or text automatically using ML or NLP techniques. Thus, inaccurate hashtags used to manipulate search results on social media platforms can be flagged by automatically by adding such hashtags using bots. Of course, end users acting manually out of a sense of civic duty could also add high quality labels to distinguish reliable information from manipulated ones.

\subsubsection{Search engines} The emergence of open-source AI models, especially open-source ones, which are connected to the latest Web results, makes it possible to retrieve information about censored topics even if the only available search engines are censored. Even though some AI models may have been finetuned to refuse to provide information directly on sensitive political events~\cite{ahmed_analysis_2025}, retrieving information using carefully constructed prompts that do not contain sensitive keywords may still be possible. More research is needed on how such prompts can be constructed reliably.

\subsubsection{App marketplaces} Technically capable users may be able to install apps directly from a trusted third-party developer or use an open-source phone OS alternative which allows users to install privacy-friendly open-source apps. 

\subsubsection{AI models} By comparing the answers generated by ChatGPT with those generated by humans, Vahid Ghafouri \textsl{et al}~\cite{ghafouri_ai_2023} demonstrated that it is difficult to completely mitigate bias from AI responses due to moderation or biased training data, despite modern AI systems having been continuously iterating and improving and trying to be neutral when asked about controversial topics. Nevertheless, some recently emerged tools, such as Ollama\footnote{https://ollama.com}, have enabled users to run various AI models (e.g., LLMs) locally by utilising container technology. Furthermore, with uncensored models contributed by the community, users can even query about sensitive topics. Since these models run in users' native containerised environments, accessing the sensitive content of the model is free from network and server-side filtering. Nevertheless, a recent study~\cite{lin_consiglieres_2025} argues that uncensored LLMs are increasingly being used to generate hate speech, violent content, or engate in other cybercriminal activities.

\subsubsection{Network throttling and monitoring} If a government tries to block some sensitive content, a decentralised \textbf{Peer-to-Peer (P2P)} network can be used as an alternative method. Users can obtain or share censored political and other sensitive content even in countries with massive surveillance and censorship infrastructure. For instance, IPFS has been adopted by communities to host content that might otherwise be subject to censorship. However, as outlined in \ref{sec:hard-res-webservers}, its content provider nodes could still undergo DoS attacks. Yet another option is to use an open-source VPN software, As noted in \ref{sec:throttling}, VPN software could be compromised, so care needs to be exercised, such as using \textit{audited} cross-platform server and client software with industry-level secure data connections.

With the advances in space technologies, satellite-powered broadband services has been deployed in many countries in recent years to help users access the Internet from remote places or restricted locations. Starlink is one example; thousands of small satellites have been, and more will be deployed to provide users with high-speed communication solutions. However, Starlink has been banned in countries like \region{China} since this could also be used to circumvent censorship~\cite{emmanuel_china_2022}.


\subsubsection{Global Monitoring} Detecting censorship and measuring its impact worldwide, especially in countries and regions under repressive regimes, are crucial to protecting human rights, such as freedom of speech. It is also beneficial for understanding how, when, and where the prevalent censorship systems operate. Systems responsible for detecting and measuring censorship usually use vantage points (probes) to collect traffic data from the censored network and compare it against the ground truth data from trusted sources. Vantage points can generally collect data from the client-side, gateway or server-side passively or actively~\cite{aceto_internet_2015}.

Global monitoring is closely connected to censorship data analysis. Both are about shedding light on censorship in the hopes that this will either shame the censoring parties into not censoring or alert users that they are being censored and that they can take steps to circumvent censorship. Data analysis is post-facto, whereas monitoring is real-time.
Arguably, measuring censorship at a global scale has always been challenging due to technical, political and ethical reasons in deploying vantage points. Some early efforts started a decade ago. In 2011, Athanasopoulos \textsl{et al.} proposed CensMon~\cite{athanasopoulos_censmon_2011}, a novel method to monitor censorship via a distributed agent network which takes probing requests from a central server and logs data for systematic analysis. Monitoring systems have evolved in the last decade. For instance, Niaki \textsl{et al.} proposed ICLab~\cite{niaki_iclab_2020} to measure a broad range of network interference and filtering in over 60 countries, with the ability to detect DNS manipulation, packet drops, and censored webpages. Around the same time, Raman \textsl{et al.} proposed Censored Planet~\cite{ram_sundara_raman_censored_2020}, a platform that collects censorship measurements on six Internet protocols (IP, DNS, HTTP, HTTPS, Echo and Discard) across over 200 countries, which complements similar platforms such as OONI\footnote{https://ooni.org}, ICLab~\cite{niaki_iclab_2020} with enhanced coverage, scalability and continuity. Raman \textsl{et al.}~\cite{raman_measuring_2020} argue that censorship events can now be detected automatically with their longitudinal measurements.

\section{Conclusion}
In some countries (and often during some periods selectively), accessing online information has become increasingly challenging. The Internet keeps evolving everyday, and accordingly, new perspectives to study Internet censorship are also emerging. In this paper, we proposed a novel taxonomy based on chokepoints, identifying where modern filtering practices occur. We defined and studied new and emerging hard and soft chokepoints; we also compared the main characteristics of these two types of chokepoints. We believe we are the first to study censorship mechanisms from this new perspective. To the best of our knowledge, our survey is the first to look at censorship practices observed at both established chokepoints, as well as new and emerging censorship chokepoints ones such as attention honeypots and shadow banning, which are becoming increasingly popular. We show increasing occurrences of Internet censorship in numerous countries, from
surveillance found on the OS level to instant messengers used by a large user base. In the context of freedom of speech being increasingly challenged in various countries, we ask for further research on emerging censorship practices gaining traction around new and emerging chokepoints, such as attention honeypots, shadow banning, and information control in AI-powered apps. 

This shift towards soft chokepoints, which are harder to detect and identify, makes development of censorship resistance systems even more challenging. For example, with attention honeypots, accounts may be operated by professional operatives such as state actors, and could look highly authentic, only acting suspiciously during sensitive political events such as elections~\cite{zannettou_disinformation_2019}. Similarly, in the case of shadow banning, users' activities or comments may be selectively suppressed or not `boosted' by the `algorithm' of a social platform. This is hard to detect or prove as the algorithms used by such platforms are often opaque. New methods, such as measuring the reach of a message from multiple network nodes in the network, need to be developed.

\bibliography{internet-censorship}

\end{document}